\documentclass[preprint,12pt]{elsarticle}

\usepackage{amssymb}
\usepackage{amsmath}
\usepackage{hyperref}
\usepackage{array}

\usepackage{changes}
\date{March 29, 2025}

\journal{BioSystems}

\begin{document}

\begin{frontmatter}



\title{Quantum information theoretic approach to the hard problem of consciousness}


\author{Danko D. Georgiev} 

\affiliation{organization={Department of Biochemistry, Molecular Medicine and Nutrigenomics, Medical University of Varna},
            addressline={84B Tzar Osvoboditel Blvd}, 
            city={Varna},
            postcode={9000}, 
            country={Bulgaria}}
\ead{danko.georgiev@mu-varna.bg}
\begin{abstract}
Functional theories of consciousness, based on emergence of conscious experiences from the execution of a particular function by an insentient brain, face the hard problem of consciousness of explaining why the insentient brain should produce any conscious experiences at all. This problem is exacerbated by the determinism characterizing the laws of classical physics, due to the resulting lack of causal potency of the emergent consciousness, which is not present already as a physical quantity in the deterministic equations of motion of the brain. Here, we present a quantum information theoretic approach to the hard problem of consciousness that avoids all of the drawbacks of emergence. This is achieved through reductive identification of first-person subjective conscious states with unobservable quantum state vectors in the brain, whereas the anatomically observable brain is viewed as a third-person objective construct created by classical bits of information obtained during the measurement of a subset of commuting quantum brain observables by the environment. Quantum resource theory further implies that the quantum features of consciousness granted by quantum no-go theorems cannot be replicated by any classical physical device.
\end{abstract}




\end{frontmatter}



\section{\label{sec:1}Introduction}

We are sentient beings who access the surrounding world only through
our conscious experiences \cite{Georgiev2017}. Without conscious
experiences, we cannot contemplate where we are, who we are and what
we are. Therefore, it is natural to identify ourselves with our \emph{conscious
minds} and define \emph{consciousness} as the subjective, private,
phenomenal, first-person point of view of our mental states, experiences
or feelings \cite{Nagel1974,Nagel1987}. To state that we are conscious minds is just a concise
way to affirm that we are sentient beings that possess inner feelings
and experiences. In other words, the \emph{mind} is equivalent to
\emph{conscious experience}. The utility of this choice of terminology
is to make the phrase \emph{conscious mind} logically redundant, since
\emph{unconscious entities} should not be called minds \cite{Georgiev2017}.
Hereafter, we will use the individual words \emph{experience}, \emph{consciousness}
and \emph{mind} interchangeably, with the understanding that \emph{unconscious
minds} or \emph{unconscious experiences} do not exist, because such
phrases translate to self-contradictory combinations of words like
\emph{unconscious consciousness}, \emph{nonmental minds} or \emph{inexperienceable
experiences}.

Having clarified what we understand under \emph{conscious mind}, we
are ready to discuss the anatomical \emph{brain}, which is composed
from individual nerve cells called \emph{neurons} that are organized
into electrically active \emph{neural networks} \cite{Yuste2015,Hahn2019,Dorkenwald2024,Kobayashi2024}.
The brain is a part of the central nervous system that appears to
be intimately related to the conscious mind because injuries to the
brain lead to cognitive deficits \cite{Rochoux1814,Walusinski2017,Ropper2019}.
Since the postmortem brain remains available for pathoanatomical examination
after the death of a person \cite{Hade2024,Park2024,Cajal}, this fact is commonly
used to support the claim that it is the living brain that produces
the conscious mind, or conversely, that the conscious mind arises
from a living brain, but not from a dead brain. Computational neuroscience
further explains how the electrically active brain neurons accomplish
the input, processing, storage, and output bits of classical Shannon information
\cite{Shannon1948a,Shannon1948b} through concerted opening and closing
of voltage-gated ion channels incorporated in the neuronal plasma
membranes \cite{Johnston1995,Flood2019,Kariev2021}.
There are physical limits on natural computation due to the existence
of the minimum action given by the Planck constant \cite{Liberman2022,Igamberdiev2025}.
This sets the boundaries to the physical world and its observation, including reflexive self-awareness in sense perception, through which the codes signifying sensual perceptive events operate and constrain human behavior \cite{Igamberdiev2024}.
Although remarkable, all computational tasks accomplished by the brain are considered \emph{easy problems}
of consciousness \cite{Chalmers1995}. According to David Chalmers,
the really \emph{hard problem} of consciousness is to explain why
the brain neurons \emph{generate} any conscious experiences at all
\cite{Chalmers1995}. In other words, why the brain does not operate
always in an insentient brain mode in which the neurons perform their
computational functions without generation of any conscious experiences?

Here, we will present a quantum information theoretic reductive solution
to the hard problem of consciousness according to which sentience
and conscious experiences are fundamental ingredients of the quantum
physical reality \cite{Georgiev2024}. Utilizing the quantum physical
laws and the formalism of quantum information theory, we will explain
how the \emph{observable brain} is created through sequential \emph{mind
choices} \cite{Georgiev2017}. In the quantum reductive approach,
the \emph{observable brain} does not produce the mind, but rather
represents \emph{what the conscious mind looks like} \emph{to external
observers} who may rely on physical measuring devices to capture and
record the observable brain data in the form of images or electric signals \cite{Georgiev2020a}.
Our approach utilizes the quantum methodology and formalism, which is also employed in recent quantum-like modeling of information processing in macroscopic neuronal structures \cite{Khrennikov2010,Khrennikov2015,Khrennikov2023}. The emphasis in the present work, however, is put on those quantum features that cannot be replicated by classical systems due to existing quantum no-go theorems \cite{Georgiev2017}. Quantum resource theory classifies such quantum features as ``genuine quantum resources'' and provides means for their faithful quantification attributing zero amount of genuine quantum resources to classical physical systems \cite{Chitambar2019,Takagi2019,Milz2020,Tendick2022}.

In order to make the exposition self-contained, in Section~\ref{sec:2} we will briefly
review the main differences between \emph{functionalism} and \emph{reductionism}
as theoretical approaches to the mind--brain problem.
Next, in Section~\ref{sec:3} we will compare the physical principles that respectively
characterize classical physics and quantum physics.
Then, in Section~\ref{sec:4} we will elaborate on the obstacles met when attempting to
address the hard problem of consciousness in the framework of classical physics.
Finally, in Section~\ref{sec:5} we will demonstrate how the hard
problem of consciousness is solved with the tools provided by quantum information theory.
We will conclude this work with a summary of the implications
of the quantum reductive approach for the evolution of consciousness,
contemplation of the consequences of our free choices, and the origin
of moral responsibility.

\section{\label{sec:2}Comparison of functionalism and reductionism}

The mind--brain problem could be approached in one of two distinct
theoretical physical ways differentiated by the direct \emph{presence}
or \emph{absence} of \emph{conscious experiences} in the physical
equations that describe the physical reality (Fig.~\ref{fig:1}).
The subsequent definitions of functionalism and reductionism aim to
capture the main difference between the two approaches that is most
relevant to the hard problem of consciousness. Different varieties
of functionalism or reductionism will be discussed later in Sections~\ref{sec:4} and \ref{sec:5} after we recall the fundamental principles of classical physics and quantum physics.

\subsection{Functionalism}

\emph{Functionalism} states that conscious experiences are produced
by the brain, which is comprised of insentient physical particles
whose motion is governed at all times by physical equations (Fig.~\ref{fig:1}A).
The characteristic feature of functionalism is that conscious experiences are not present inside the physical equations, but it is possible to \emph{turn} emergent conscious experiences \emph{on} or \emph{off}
depending on the circumstances, e.g., a healthy living brain
is considered capable of producing conscious experiences in its awake
state, whereas no conscious experiences are produced during brain
concussion, sudden drop of blood glucose levels or application of
general anesthetics \cite{Georgiev2017}.

Functionalism is based on the premise of \emph{naive realism} according
to which ``what is observed'' is necessarily ``what exists'' in
the physical reality. Since the anatomical brain is what is observed
while peeking inside the skull of a patient undergoing open skull
neurosurgery \cite{Penfield1978}, it is assumed that it is the brain
that physically exists, whereas the first-person, private, unobservable,
phenomenal conscious experiences are somehow produced by the brain
\cite{Fodor1981}. The hard problem of consciousness then is to explain
why the brain is obliged to produce any conscious experiences rather
than remaining insentient or unconscious, devoid of any feelings,
while still functioning. In fact, the hard problem is a characteristic
feature of functionalism, which does not arise in reductionism, as
we shall see next.

\subsection{Reductionism}

\begin{figure}[t]
\begin{centering}
\includegraphics[width=\textwidth]{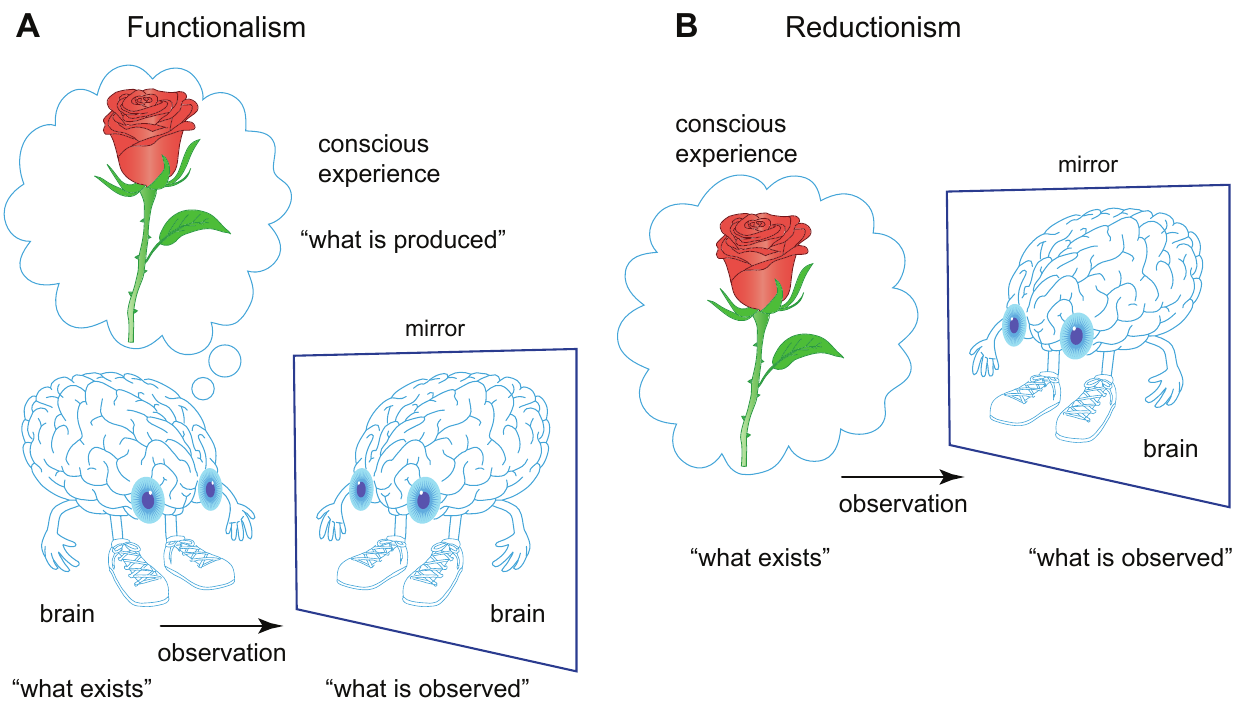}
\par\end{centering}

\caption{\label{fig:1}Two alternative theoretical approaches to the mind--brain
problem. (A) Functionalism states that the anatomical brain produces
conscious experiences. The brain is both ``what exists'' in the
physical reality and ``what is observed'' by external observers
or physical measuring devices. (B) Reductionism states that the conscious
experience is ``what exists'' in the physical reality, whereas the
brain is ``what is observed'' by external observers or physical
measuring devices. The brain does not produce the conscious experience,
but rather acts as a ``label'' that signifies the existing conscious
experience. Legend: the red rose represents phenomenal conscious experience,
the brain cartoon represents the anatomical brain, and the mirror
represents what can be observed by external observers or physical
measuring devices.}
\end{figure}

\emph{Reductionism} states that conscious experiences are already
present inside the physical equations that govern the temporal dynamics
of physical systems (Fig.~\ref{fig:1}B). This means that the mind
is identified with the constituent physical particles, which always
remain sentient and their conscious experiences can never be turned
off. In the reductive approach, the brain is just a label that refers
to the conscious experiences that exist in the physical reality. The
label ``brain'' is neither better nor worse when compared to the
phrase ``conscious experiences'' that also refers to the existing
conscious experiences.

Reductionism denies the veracity of \emph{naive realism} and affirms
the opposite, namely, that conscious experience is ``what exists''
physically, whereas the brain is what the conscious experience looks
like to external observers or physical measuring devices. Because
the brain is only ``what is observed'' but has no separate fundamental
existence, it cannot produce the conscious experiences. To the contrary,
the existing conscious experiences are the physical entities that
produce the observable brain when subjected to physical measurement
\cite{Georgiev2020a}. The hard problem of consciousness is avoided
entirely and never arises in reductionism because one cannot meaningfully
ask why the brain produces conscious experiences since the reductive
approach denies that the brain produces any conscious experiences.
If anything, it is the other way around, namely, the existing conscious
experiences produce the observable brain, because the conscious experiences
have to exist first in order to \emph{look like something} to an external
observer.

In reductionism, one may ask the converse question, namely, why does
the conscious mind look like a brain from an external, third-person
point of view, however, the answer to this kind of question is quite
easy: The conscious mind should look like ``something'' and the
actual shape and texture of the brain is just an \emph{evolutionary
accident} due to evolution through natural selection of organisms
that have bilateral symmetry \cite{DeFelipe2011,Marino2007,Herculano2009,Herculano2014}.
The fact that the conscious mind should look like ``something''
to the external world is a consequence from the requirement that the
mind \emph{interacts with} and \emph{has some physical effect upon}
the surrounding world \cite{Georgiev2024}.

\section{\label{sec:3}Comparison of classical mechanics and quantum mechanics}

Before we focus on the merits of the quantum reductive solution to
the hard problem of consciousness, it is worth recalling the main
principles characterizing classical physics and quantum physics. For our present
purposes, it suffices to limit ourselves to the nonrelativistic domain
in order to compare Hamiltonian classical mechanics with Hamiltonian
quantum mechanics. Having an explicit list of all the important differences
between classical mechanics and quantum mechanics will be extremely helpful
for the analysis of illustrative examples on the mind--brain relationship.

\subsection{\label{sub:31}Classical mechanics}

Classical mechanics can be formulated in several different, but mathematically
equivalent ways using Newtonian, Lagrangian or Hamiltonian formalism
\cite{Susskind1}. The Hamiltonian formalism, which relates time-independent
Hamiltonians with the total energy of the system (see \ref{app1}), is best suited for
comparison with nonrelativistic quantum mechanics due to certain similarities
between the mathematical structures that arise from Dirac's explicit
quantization rule that replaces classical Poisson brackets with quantum
commutators \cite{Ashtekar1980}. The three main characteristics of
the classical physical world can be summarized in the form of fundamental
principles or postulates:

\paragraph{C1 (classical states are observable)}

The classical physical state of each classical system is an \emph{observable}
that can be represented as a point in multidimensional phase space
in which every physical degree of freedom is represented as an axis.

\paragraph{C2 (temporal dynamics is deterministic)}

The temporal dynamics of each classical physical state is governed
by \emph{deterministic} Hamilton's equations. This means that the
dynamics of the classical world resembles the working of a ticking
clockwork mechanism whose future behavior is completely determined
by the physical laws and the initial physical state of the universe.

\paragraph{C3 (unobservable things do not exist)}

Everything that exists in the classical physical world is \emph{observable}
and the existence of things can be verified using physical measurements
with measuring devices. Logically, this also implies that \emph{unobservable}
things do not exist in the classical world.\\

Although the principles of classical physics may appear innocent,
we will see in Section~\ref{sec:4} how they actually lead to insurmountable
paradoxes in the theory of consciousness.

\subsection{\label{sub:32}Quantum mechanics}

Quantum physics is the modern, most successful, experimentally verified
description of the physical world we live in \cite{Susskind2}. Its
radical departure from classical notions is manifested in the fact
that ``what exists'' in the quantum reality is different from ``what
can be observed''. The latter difference is expressed in using different
mathematical objects for \emph{quantum states} and \emph{quantum observables}
\cite{Dirac1967,vonNeumann1955} (see \ref{app2}). The three main characteristics of
the quantum physical world can be summarized in the form of fundamental
principles or postulates:

\paragraph{Q1 (quantum states are unobservable)}

The quantum physical state $\vert\psi\rangle$ of an isolated quantum
system can be represented as a vector in $n$-dimensional complex
Hilbert space $\mathcal{H}$. The quantum state vector $\vert\psi\rangle$
is \emph{unobservable}. Mathematically, the quantum state vector $\vert\psi\rangle$
could be viewed as a $n\times1$ matrix when expressed in a given
basis of the Hilbert space $\mathcal{H}$.

\paragraph{Q2 (temporal dynamics is indeterministic)}

The unitary temporal dynamics of the quantum state vector $\vert\psi\rangle$
of an isolated quantum system is governed by the Schrödinger equation.
The temporal dynamics of the quantum state vector $\vert\psi\rangle$
can be used to predict only the quantum probabilities for different
potential measurement outcomes according to the Born rule, however,
in general it cannot determine the exact measurement outcome that
will be actualized upon measurement of the quantum system. Thus, individual
quantum measurement outcomes are in general \emph{indeterministic}.

\paragraph{Q3 (both unobservable and observable things do exist)}

Only quantum \emph{observables} $\hat{A}_{i}$ are subject to observation
in the quantum physical world. The quantum observables $\hat{A}_{i}$
are operators on the Hilbert space $\mathcal{H}$ and their action
could be understood as changing input quantum state vectors to output
quantum state vectors, when an initially isolated quantum system is
subjected to physical measurement with external measuring devices.
Mathematically, the quantum observables $\hat{A}_{i}$ could be viewed
as $n\times n$ matrices when expressed in a given basis of the Hilbert
space $\mathcal{H}$. The quantum probabilities for different measurement
outcomes depend both on the quantum observable $\hat{A}_{i}$ and
the unobservable quantum state vector $\vert\psi\rangle$ according
to the Born rule. The actualized measurement outcome does not preexist,
but is generated at the time of quantum measurement, which follows
from the Kochen--Specker theorem \cite{Kochen1967,Georgiev2017}.\\

Since not everything that exists in the quantum physical world is
observable, it is no longer possible to just input the initial quantum
state of the universe and let it evolve unitarily in time. The introduction
of intermittent quantum measurements leads to indeterministic quantum
jumps, which leave the future evolution of the universe open to the
actual choices performed by the constituent quantum systems whenever
they are subjected to measurements.

In order to be able to perform concrete calculations and make genuine
quantum predictions, one would need to master a mathematically precise
list of quantum axioms and rules for implementation of those axioms
\cite[\S4]{Georgiev2017}. For the present purposes, however, we will
work with the quantum principles at an abstract conceptual level that
highlights the difference between quantum state vectors and quantum
observables, and we will provide quantum bra-ket notation only to
facilitate background reference to specialized literature \cite{Dirac1967,vonNeumann1955,Hayashi2015,Griffiths2018}.

\section{\label{sec:4}The hard problem in classical physics}

\subsection{\label{sub:41}Classical functionalism}

Classical functionalism states that the neural networks in the brain
produce conscious experiences. Different varieties of classical functionalism
propose different computational ``functions'', which if performed
as prescribed, will generate conscious experiences. Taking into account
the mathematical structure of classical physics presented in Section~\ref{sub:31}, however, it could be proved as a theorem that any conscious
experience generated by the brain would necessarily be a causally
impotent epiphenomenon that cannot affect anything in the physical
world, hence the production of conscious experiences by the brain
would be indistinguishable from lack of production of any conscious
experiences \cite{Georgiev2023}. The proof of the latter theorem
goes as follows:

First, it is noticed that the deterministic physical laws are obeyed
by all classical physical systems, including the brain. This means
that the temporal dynamics of the brain has to follow exactly the
Hamilton's equations \cite{Hamilton1833} and it is not up to the
brain to decide to obey or disobey the classical physical laws. The
mathematical properties of Hamilton's equations, which are deterministic
differential equations, will then ensure that what the brain does
in the future is only dependent on the initial insentient brain state (including some fixed initial state of the environment, which we can ignore without loss of generality).
If the initial insentient brain state is such that the future brain
dynamics executes the computational ``function'' required to generate
conscious experiences, these conscious experiences will be generated
but they will not be able to affect in any way the temporal dynamics
of the brain, which is already completely determined by the initial
insentient brain state and the Hamilton's equations. To convince oneself
that the generated conscious experiences are causally impotent with
regard to the brain dynamics, one can subtract the computed brain
dynamics in case 1 when the initial insentient brain state and the
Hamilton's equations produce conscious experiences and case 2 when
the initial insentient brain state and the Hamilton's equations do
not produce any conscious experiences. The result from the subtraction
of case 2 from case 1 will always be zero because solving differential
equations given initial state does not care about mathematically irrelevant
stuff such as whether conscious experiences are produced in the process.
This problem of causally impotent consciousness is not new and has
been referred to as the problem of \emph{mental causation} in philosophical
literature \cite{Kim1998}. Here and in previous works \cite{Georgiev2017,Georgiev2023},
we have only emphasized that the problem of mental causation is not
pervasive in all of natural science, but only limited to the particular
combination of \emph{functionalism} and \emph{determinism} in \emph{classical
physics}.

The hard problem of consciousness, contained in the question why the
brain produces any conscious experiences at all, is exacerbated by
classical functionalism because any conscious experiences produced
by the brain are necessarily causally impotent in the physical world.
The computational ``functions'' performed by the brain could be
useful, while the accompanying conscious experiences or feelings would
be utterly useless. Then, the advocates of functionalism have to explain
not only why the brain produces something that is not necessitated
by the physical laws, but also why the brain produces something that
is utterly useless. The most promising way out of the impasse is to
reject functionalism and investigate how reductionism handles the
mind--brain problem.

\subsection{\label{sub:42}Classical reductionism}

\begin{figure}[t]
\begin{centering}
\includegraphics[width=\textwidth]{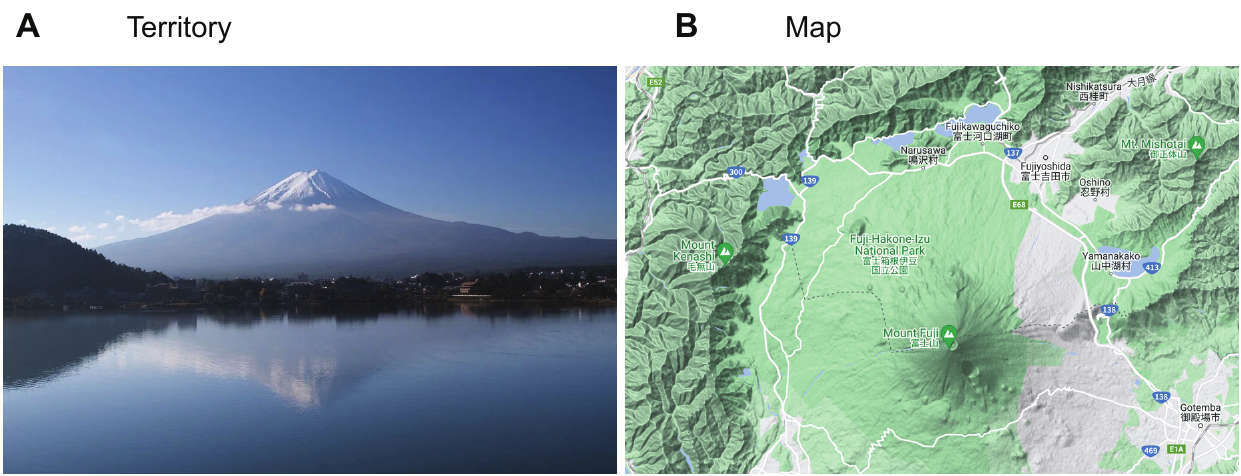}
\par\end{centering}

\caption{\label{fig:2}The map is not the territory. (A) Photograph of Mount
Fuji reflected on Lake Kawaguchi, near Fujikawaguchiko town, Yamanashi
prefecture, Japan. The photograph represents the territory that exists
in the physical reality. (B) Relief map portraying the elevation of
Mount Fuji and its geographical location with respect to Fujikawaguchiko
town on the shore of Lake Kawaguchi. Although the map is distinct
from the territory, the map represents communicable knowledge about
the territory.}
\end{figure}

Reductionism avoids completely the hard problem by introducing conscious
experiences directly into the physical equations that describe the
physical reality. This is achieved by identification of the mind and
the brain within the scientific theory (i.e., both the mind and the brain \emph{refer to} the same physical reality), thereby implying that the initial brain state is already
sentient and no further production of conscious experiences is needed.
The reductive approach works by highlighting the distinction between
the \emph{map} and the \emph{territory} \cite{Korzybski1994}. The conscious experiences,
which exist in the physical reality, are like the territory of Mount Fuji (Fig.~\ref{fig:2}A), whereas the brain is like the map of Mount Fuji (Fig.~\ref{fig:2}B), which refers to the same physical reality (Mount Fuji)
by capturing some of its properties, such as geographic localization,
in the form of communicable classical Shannon information.

The possibility of having several words referring to the same physical
reality was already exemplified by our convention to use the words
``experience'', ``consciousness'' and ``mind'' interchangeably
\cite{Georgiev2017}. In this respect, adding the word ``brain''
as another label for the existing conscious experiences is not something
that is conceptually challenging to understand. What is interesting,
however, is the realization that besides using single word labels,
one is also able to use longer descriptions to refer to physical reality,
similarly to how the map captures existing properties of the territory.
For example, rather than just saying ``pain'', it is possible to
refer to the same painful experience by describing the ``activation
of certain neurons in the brain''. Conceptually, the implication
is that the ``activation of certain neurons in the brain'' is not
producing the pain experience, but rather is ``what the pain experience
looks like to external observers''. When generalized beyond the particular
type of conscious experience, reductionism affirms that the mind is
what exists, whereas the brain is ``what the mind looks like to external
observers'' (Fig.~\ref{fig:1}B).

The main advantage of reductionism over functionalism is the absence
of the hard problem of consciousness. This can be illustrated by the
following comparison:

In functionalism, the brain is comprised of insentient physical particles ($\approx~80\%$ water),
which move around in space and miraculously produce conscious experiences.
Suppose for example that a water molecule in the brain moves consecutively
between three positions in space, where the motion is denoted as $1\to2\to3$,
given the premise that both the water molecule is insentient and the
physical position space is insentient. Then, the hard problem is to
explain: why the motion $1\to2\to3$ is obliged to produce any conscious
experiences at all, and how the motion $1\to2\to3$ that produces
conscious experiences is distinguishable from the motion $1\to2\to3$
which does not produce any conscious experiences. If the answer is
that the motion $1\to2\to3$ \emph{is the same} (based on Hamilton's
equations) \emph{with} or \emph{without} the production of conscious
experiences, then the produced conscious experiences are an epiphenomenon
that lacks any causal potency upon the physical motion in space and
the presence of conscious experiences is indistinguishable from their
absence in the physical world \cite{Georgiev2023}.

In reductionism, the brain is comprised of sentient physical particles,
which move inside a sentient, mental space. In this case, the word
``brain'' denotes the mind, and positions in the mental space denote
feelings or distinct conscious experiences. For example, 1 can denote
``happy'', 2 can denote ``surprised'', and 3 can denote ``sad''.
Mathematically, the motion in mental space can be denoted again as
$1\to2\to3$, however, now the meaning is an alteration of the conscious
experience from ``happy'' to ``surprised'' and then to ``sad''.
The motion in mental space does not produce conscious experiences,
but rather represents the \emph{stream of conscious experiences} changing
in time. There is no hard problem of consciousness because no production
of conscious experiences is required in a mental space. Similarly
to how the map is not producing the territory, if the physical reality
is fundamentally mental, there is no need for the conscious mind to
be produced by the brain. The existing mind is like the \emph{territory},
whereas the observable brain is like the \emph{map} that conveys communicable bits of
classical Shannon information about the territory \cite{Georgiev2020a}.

\begin{table}
\caption{\label{tab:1}Comparison of functionalism and reductionism in classical
physics.}

\begin{tabular}{|>{\centering}p{0.5\textwidth}|>{\centering}p{0.5\textwidth}|}
\hline 
\textbf{Classical functionalism} & \textbf{Classical reductionism}\tabularnewline
\hline 
\hline 
The brain is comprised of insentient particles. & The brain is comprised of sentient particles.\tabularnewline
\hline 
The brain is not the conscious experience. & The brain is the conscious experience.\tabularnewline
\hline 
The brain produces the conscious experience. & The conscious experience \emph{just is} what it is.\tabularnewline
\hline 
The conscious experience is causally ineffective. & The conscious experience is causally effective.\tabularnewline
\hline 
The hard problem is to explain why we need any conscious experiences
at all. & There is no hard problem because conscious experience is fundamental
ingredient of reality.\tabularnewline
\hline 
\end{tabular}

\end{table}

Although, at this point the superiority of reductionism over functionalism
appears to be established (Table \ref{tab:1}), it is worth noting
that quantum physics is able to provide a great number of additional
advantages over classical physics. For example, classical reductionism
has a number of deficiencies that result directly from the fundamental
principles of classical physics:

First, conscious experiences are not directly observable by external
observers, whereas one of the fundamental principles of classical
physics states that everything that exists in the classical physical
world is observable. Employing \emph{modus tollens} as a rule of inference \cite[p.~43]{Georgiev2017},
we can conclude that everything that is not observable does not exist
in the classical physical world. The contradiction is that conscious
experiences do exist, whereas the classical physics implies that if
conscious experiences are not observable they cannot exist. The solution
is to reject classical physics as incorrect theory and adopt quantum
physics, which contains mathematically precise description of physical
things that are unobservable \cite{Georgiev2020b}.

Second, the phenomenological character of conscious experiences referred
to as \emph{qualia}, such as the ``blueness'' of the blue sky, the
``painfulness'' of a hand injury, or the ``sonar image'' seen
by a bat, cannot be communicated to others in the form of classical
bits of Shannon information. The existence of non-communicable things
in classical physics is exactly as problematic as the existence of
unobservable things, since everything that exists in the classical
world is communicable. Again, employing \emph{modus tollens}, we can conclude that non-communicable things should not exist. Similarly, the solution is to reject classical physics as incorrect
theory.

Third, introspectively we feel that we possess free will manifested
in our genuine ability to make choices for future courses of action,
whereas another of the fundamental principles of classical physics
states that the temporal dynamics of all existing physical things
is deterministic. The problem of lacking free will is not only that
our introspective testimony of possessing free will would be untrustworthy,
but also that it would follow that we are not morally responsible
for our actions because we could not have done otherwise \cite{Georgiev2021}.
Here again, the solution is to reject classical physics as incorrect
theory and adopt quantum physics, which provides indeterministic dynamics
and allows for quantum agents to make genuine choices for future courses
of action \cite{Georgiev2017,Georgiev2023,Georgiev2021}.

\section{\label{sec:5}The hard problem in quantum physics}

\subsection{Quantum functionalism}

The introduction of quantum physics in the theory of consciousness
would be useless if one insists on keeping functionalism. Indeed,
postulating that quantum particles are insentient, while at certain
\emph{special} circumstances the quantum particles generate conscious
experiences, would bring us back to the hard problem, which would
require an explanation of why any conscious experiences at all are
generated under those \emph{special} circumstances. The hard problem
will be a little bit different in quantum functionalism compared to
classical functionalism because the temporal dynamics of quantum systems
is indeterministic. The indeterminism means that mathematically repeating
the stochastic quantum dynamics with exactly the same initial conditions
will produce different outcomes for different runs \cite{Georgiev2023}.
This feature of indeterminism due to the nature of stochastic It\^{o}
calculus \cite{Ito1944,Ito1961,Gudder1979,Glimm1987} does not allow
one to prove with mathematical precision that the produced conscious
experiences will be a causally ineffective epiphenomenon, as it is the
case in classical functionalism. Instead, one could hypothesize that
the produced conscious experiences \emph{choose} the different physical
outcomes for different runs. This latter hypothesis could not be ruled
out a priori, but would have to be refuted based on introspective
experimental evidence. For instance, we could verify introspectively
that when we make our conscious choices, we do not consciously choose
which neuron will fire and which synapse will release neurotransmitter
molecules in our brain cortex \cite{Georgiev2020c}. Thus, quantum
functionalism based on emergent consciousness that selects the resulting quantum physical brain states is disproved empirically as it contradicts our daily introspective testimony.

An alternative version of quantum functionalism is provided by Hameroff--Penrose orchestrated objective reduction (Orch OR) model according to which the unitary quantum dynamics resulting in quantum coherent superposition of brain microtubules constitutes \emph{pre-conscious} processing, followed by objective reduction that produces a \emph{flash of conscious experience} \cite{Hameroff1996}. A characteristic feature of this proposal is that conscious experiences are discrete, discontinuous events occurring at a frequency of 40 Hz in-between continuous time intervals with unconscious brain activity. The main problem of this theoretical construction is that consciousness becomes a causally impotent epiphenomenon since it is a product of the objective reductions, but it is not the agent that causes the objective reductions. In fact, only if the conscious experiences are continuously present throughout the whole duration of the unitary quantum dynamics, it would be possible to model the outcomes of quantum measurements and the objective reductions as a form of conscious choice or decision making by a causally effective consciousness \cite{Georgiev2017}. This brings us to the theoretical merits of quantum reductionism, which we shall discuss next.

\subsection{Quantum reductionism}

Because quantum physics contains both \emph{observable} and \emph{unobservable}
physical entities, the reductive theory of consciousness is able to
correctly identify conscious experiences with unobservable quantum
information carried by the quantum state vector $|\psi\rangle$ of
a quantum system \cite{Georgiev2017}. Following the map vs the territory
distinction shown in Fig.~\ref{fig:2}, we could state that the conscious
experience is the real physical thing that exists as ``territory''
in the physical reality, whereas the quantum state vector $|\psi\rangle$
is the conceptual representation of the conscious experience inside
the ``map'' that is our scientific theory of the world. The resulting
quantum reductionism keeps all the advantages of the reductive approach,
which are discussed in Section~\ref{sub:42}.

\begin{figure}[t!]
\begin{centering}
\includegraphics[width=\textwidth]{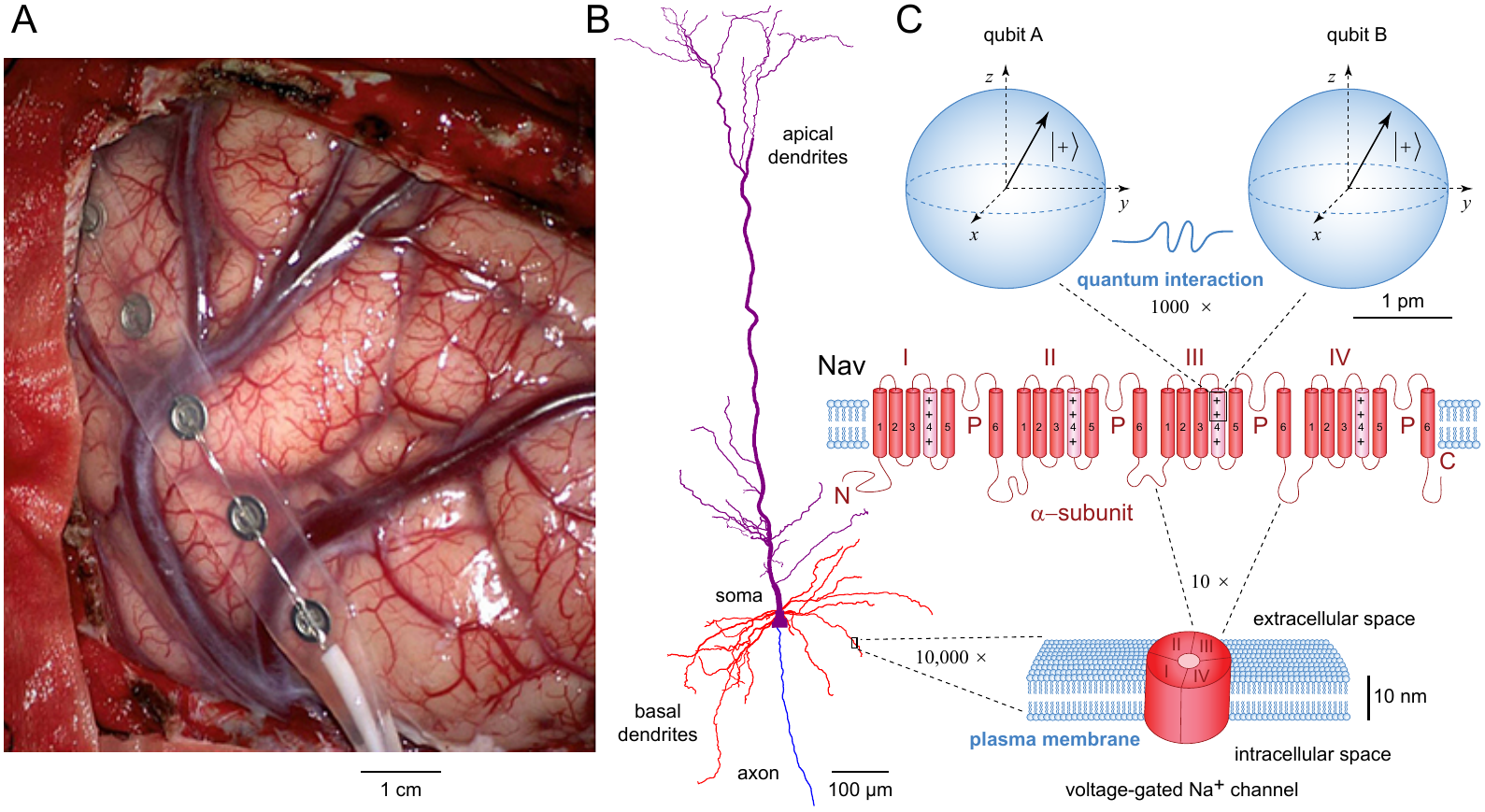}
\par\end{centering}

\caption{\label{fig:3}The observable brain cortex is what the conscious mind
looks like from a third-person point of view. (A) Photograph of the
human brain cortex obtained during open skull neurosurgery \cite{You2021}. (B) At the microscopic scale, the brain cortex is composed of neurons, which form neural networks. The morphology of pyramidal neuron (NMO\_05726) from cortical layer~5 \cite{BarYehuda2008} (http://NeuroMorpho.Org; accessed on 3 March 2025) reflects the functional specialization of neuronal dendrites and axon. (C) At the nanoscale, the electric activity of
neurons is generated by voltage-gated ion channels, which are inserted
in the neuronal plasma membrane. The magnified view shows a single voltage-gated Na$^{+}$ channel (Nav) composed of one pore-forming $\alpha$-subunit through which the Na$^{+}$ ions may pass. The $\alpha$-subunit
consists of four repeat domains, labeled I through IV, each containing
six $\alpha$-helices traversing the plasma membrane. The 4th $\alpha$-helix
is positively charged and acts as a voltage sensor \cite{Catterall2010}.
At the picoscale, individual elementary electric charges within the
protein voltage sensor could be modeled as qubits represented by Bloch
spheres. For the diameter of each qubit is used the Compton wavelength
of electron \cite{Georgiev2020c}. Consecutive magnifications from micrometer ($\mu$m)
to picometer (pm) scale are indicated by $\times$ symbol.}
\end{figure}

The brain is what the mind looks like when observed by external observers
(Fig.~\ref{fig:3}). Because in quantum physics, there are fundamentally
two kinds of physical objects, namely, quantum states and quantum
observables, when we talk about the observable brain, we specifically
invoke quantum physical observables of the brain described by quantum
operators like position $\hat{x}$ or momentum $\hat{p}$. Due to
\emph{quantum complementarity}, noncommuting quantum observables cannot
be measured or observed at the same time \cite{Georgiev2021b,Gao2023,Gudder2023,Gudder2024}.
The application of quantum measurement theory to modeling human cognition is able to explain the existence of question order effect and response replicability effect in decision making \cite{Ozawa2021}.
The incompatibility of noncommuting quantum observables implies that there is an uncountable set of potential quantum observables that could have been measured, but at a single instance of time $t$
only a single subset of commuting quantum brain observables can be actually measured.
This shows that in quantum physics, there is both a \emph{counterfactual brain} comprised of all the potential quantum brain observables that could have been measured (but were not measured)
and the \emph{observable brain} comprised of the actual quantum brain
observables that have been measured.

The physical outcomes obtained during the measurement of the observed quantum brain observables is
what or how the mind looks like from an external point of view. Why the
mind does not look the same as the first point of view is a meaningful
question that produces a paradox in classical physics, because what
exists is what is observed classically (Fig.~\ref{fig:1}). In quantum physics, what exists
is certainly not what is observed, and this is expressed neatly in
the fact that two different mathematical objects are used to describe
quantum states (vectors) and quantum observables (operators).
Why conscious experiences that exist in nature look like a collection of electrons and protons from a third point of view, forming biomolecules in the human brain, can be answered as a fact of natural evolution.
Quantum physics forbids the possibility for the conscious experiences to look like conscious experiences from a third point of view and this physical fact explains neatly the inner privacy of consciousness.
The inner privacy of consciousness encompasses two closely linked, yet distinct, physical concepts: \emph{unobservability} (the characteristic of defying observation or measurement) and \emph{incommunicability} (the characteristic of not being communicable through classical bits of information such as words). The interplay between these two concepts can be articulated as follows: if conscious experiences were observable, we would have possessed the capacity to convey the phenomenal essence of such experiences merely by allowing others to observe us; conversely, if conscious experiences were communicable, we would have had the means to recover missing senses solely through verbal expression \cite{Georgiev2020b}.
Furthermore, only the quantum reductive approach is consistent with our introspective
testimony that we choose conscious experiences as outcomes, e.g.,
conscious decision of moving one's own arm, rather than consciously
choosing the observable brain outcomes, e.g., which neuron in the
motor brain cortex will fire electric action potential in order to
trigger muscle contraction in one's own arm (Table~\ref{tab:2}).

The quantum reductive approach to consciousness has experimental implications, which stem from the identification of mental states with quantum entangled state vectors in the anatomical brain. This requires underlying biomolecular substrates in the brain that exhibit quantum behavior as prescribed by the Schr\"{o}dinger equation. Recent advances in computational quantum chemistry has provided evidence for dynamic quantum effects in enzyme catalysis \cite{Alhambra1999,Luk2015,Wang2016,Yang2019}, ion channel gating \cite{Kariev2021}, protein dynamics \cite{Georgiev2022x} and exocytosis of synaptic vesicles \cite{Georgiev2018x}. Quantum tunneling in the gating of voltage-gated ion channels or synaptic exocytosis of neurotransmitters may act as a quantum trigger whose effects are amplified into macroscopic patterns of electric activity of the cortical neural network. Experimental confirmation of genuine quantum effects inside the voltage-sensors, selectivity filters and pores of ion channels \cite{Vaziri2018,Wang2024,Georgiev2021x} would render purely classical explanation of consciousness unlikely and would strengthen the quantum information theoretic approach.

\begin{table}
\caption{\label{tab:2}Comparison of functionalism and reductionism in quantum
physics.}

\begin{tabular}{|>{\centering}p{0.5\textwidth}|>{\centering}p{0.5\textwidth}|}
\hline 
\textbf{Quantum functionalism} & \textbf{Quantum reductionism}\tabularnewline
\hline 
\hline 
The quantum brain is comprised of insentient particles. & The quantum brain is comprised of sentient particles.\tabularnewline
\hline 
The quantum brain is not the conscious experience. & The quantum state $|\Psi\rangle$ of the brain is the conscious experience.\tabularnewline
\hline 
The quantum brain produces the conscious experience. & The conscious experience \emph{just is} what it is.\tabularnewline
\hline 
The conscious experience chooses the future quantum brain state. & The conscious experience chooses the outcome for future conscious
experiences.\tabularnewline
\hline 
The theory is introspectively incorrect because we \emph{do not know}
what our quantum brain state is, hence we could not have chosen it. & The theory is introspectively correct because \emph{being} what we
are does not imply \emph{knowing} what we are.\tabularnewline
\hline 
\end{tabular}
\end{table}

\section{\label{sec:6}Conclusion}

Functional theories of consciousness do not simply state that consciousness
is \emph{what brains do} because during general anesthesia the human brain seems
to temporarily lose consciousness. However, insisting that conscious
experiences can be \emph{turned on} or \emph{off}, depending on what
kind of function the brain performs, leads to immediate clash with
the mathematical properties of \emph{differential equations} that govern
the temporal dynamics of physical systems. In particular, the temporal
dynamics resulting from deterministic ordinary differential equations
is insensitive to any postulated emergence of conscious experiences,
which means that emergent conscious experiences cannot affect the
physical behavior of organisms in any way and would be an evolutionary
useless epiphenomenon \cite{Georgiev2023,Popper1983}. In this case, the hard
problem of consciousness becomes intractable, because one is expected
to answer the question what a physically useless consciousness is
useful for?

Evolutionary biology asserts that we have conscious experiences, because
consciousness is a useful thing to have and organisms endowed with
consciousness have an edge in the struggle for survival \cite{Darwin1896,Sherwood2008}. This can
be true only within a physical framework that allows for causally
potent conscious experiences \cite{Georgiev2023}. Fortunately, modern
quantum physics provides the necessary physical tools that help the
modeling of human consciousness, cognition, creativity and thinking
as a quantum process that occurs inside the anatomical brain \cite{Melkikh2019,Melkikh2023}.
The identification of mental states, which are phenomenal states of
conscious experiences, with quantum states $|\Psi\rangle$ in the
brain, allows for immediate application of different quantum no-go
theorems to consciousness research \cite{Georgiev2013}. A detailed
list of significant results has been previously provided in Refs.~\cite{Georgiev2017,Georgiev2020a,Georgiev2023},
where the uniqueness of the human mind is protected against duplication
by the quantum \emph{no-cloning theorem} \cite{Wootters1982}, the
communicability of a certain number of classical bits of information about the contents
of conscious experience is established by \emph{Holevo's theorem}
\cite{Holevo1973}, and the causal potency of the human mind is granted
by the \emph{Born rule} \cite{Born1926,Georgiev2023,Georgiev2021}.
The hard problem of consciousness does not arise in the quantum reductive
approach because conscious experiences are fundamental ingredients
of physical reality that are attributed to all elementary quantum
particles. In this framework, the difference between an electrically
active living brain and a homogenized dead brain tissue at a thermal
equilibrium is not that the former produces conscious experiences,
while the latter lacks conscious experiences, but rather that the
living brain is what a human conscious mind looks like to an external
observer, while the homogenized dead brain is what a collection of
stochastic, random, memoryless, elementary conscious experiences looks
like to an external observer \cite{Georgiev2020a}.

At the end of 19th century, William James \cite{James1890} argued
that evolutionary psychology requires ``mind dust'' if natural evolution
is to work smoothly. His argument was correct, even though James maintained
his distaste for the conception of ``mind dust'' at a time when
classical physics was the only kind of physics known to science. Fortunately,
with the discovery of quantum physics in 1920s, we now have accumulated
knowledge about the existence and properties of \emph{complex quantum
probability amplitudes} \cite{Feynman2013}, which are an excellent
physical candidate to fulfill the roles required from a true ``mind
dust''. The quantum physicist Freeman Dyson agrees with the latter
statement eloquently: ``Our consciousness is not just a passive epiphenomenon
carried along by the chemical events in our brains, but is an active
agent forcing the molecular complexes to make choices between one
quantum state and another. In other words, mind is already inherent
in every electron, and the processes of human consciousness differ
only in degree but not in kind from the processes of choice between
quantum states which we call “chance” when they are made by electrons”
\cite{Dyson1979}.

The capacity of elementary quantum particles to perform genuine choices
for future courses of action is consistent with the attribution of
elementary conscious experiences or ``mind dust'' to all quantum physical systems.
It also vindicates the trustworthiness of our own
introspective testimony according to which we feel as morally responsible
beings that possess free will and could have done otherwise \cite{Georgiev2021,Kane2014}.
Living inside a quantum physical world built up from a fabric that
possesses mental features suggests that the origin of life inside
the universe is not a freak accident, but probably a normal occurrence
throughout the cosmos at all places that possess abundance of chemical
elements and a reliable source of free energy to sustain self-organization
\cite{Oparin1957,Darwin1859,Saha2018}.

\section*{Funding}

This study is financed by the European Union-NextGenerationEU through
the National Recovery and Resilience Plan of the Republic of Bulgaria,
project No. BG-RRP-2.004-0009-C02, Research group 3.1.1. Natura4Health.


\appendix
\section{Axioms of classical mechanics}
\label{app1}

The precise mathematical formulation of the axioms of classical mechanics can be summarized as follows \cite{Georgiev2017,Susskind1,Hamilton1833}:

\paragraph{C1 (state)}
The state of a classical mechanical system composed of $n$~particles is completely specified by a point~$(q_{i},p_{i})$ in $6n$-dimensional real phase space~$\mathbb{R}^{6n}$, where $q_{i}$ are generalized position coordinates, $p_{i}$ are generalized momentum coordinates, and the index $i=1,2,\ldots,3n$ enumerates the three spatial dimensions of each particle in the system.

\paragraph{C2 (observables)}
The observables of a classical mechanical system composed of $n$ particles are given by mathematical
functions $f(q_{i},p_{i},t)$ on the real phase space $\mathbb{R}^{6n}$.
This implies that the state of a classical mechanical system is an observable entity that is time-dependent.

\paragraph{C3 (dynamics)}
There is a distinguished observable corresponding to the total energy
of the system, called the Hamiltonian function $H(q_{i},p_{i},t)$,
that uniquely determines the time evolution of the classical physical system according
to Hamilton's equations:
\begin{equation*}
\frac{dq_{i}}{dt} =\frac{\partial H}{\partial p_{i}}, \qquad
\frac{dp_{i}}{dt} =-\frac{\partial H}{\partial q_{i}}
.
\end{equation*}
Due to the deterministic nature of Hamilton's equations, only observable physical quantities can affect the behavior of classical physical systems. This could be understood as a closure of the classical physical world towards the influence of unobservable entities, which would be classified as non-physical.

The time evolution of any classical observable $f(q_{i},p_{i},t)$ can
be calculated with the use of the total derivative in time $t$ as
\begin{equation*}
\frac{df}{dt}=\frac{\partial f}{\partial t}+\sum_{i}\left(\frac{\partial f}{\partial q_{i}}\frac{\partial H}{\partial p_{i}}-\frac{\partial f}{\partial p_{i}}\frac{\partial H}{\partial q_{i}}\right)
.
\end{equation*}

\section{Axioms of quantum mechanics}
\label{app2}

The main features of quantum mechanics could be formulated with the use of Dirac--von Neumann axioms as follows \cite{Dirac1967,vonNeumann1955,Georgiev2021x}:

\paragraph{Q1 (state)}
The quantum physical state of a closed system is fully described by a unit state vector $|\Psi\rangle$ in complex-valued Hilbert space $\mathcal{H}$.
The quantum state of a composite quantum system comprised from $k$~components lives in a tensor product Hilbert space given by the tensor product $\otimes$ of the state spaces of the component subsystems $\mathcal{H}=\mathcal{H}_{1}\otimes\mathcal{H}_{2}\otimes\ldots\otimes\mathcal{H}_{k}$.

\paragraph{Q2 (observables)}
To every observable quantum physical property $A$ there exists
an associated Hermitian operator $\hat{A}=\hat{A}^{\dagger}$, which
acts on the Hilbert space of states $\mathcal{H}$. The~eigenvalues
$\lambda_{A}$ of the operator $\hat{A}$ are the possible values
of the observable quantum physical property.
According to the Born rule, the expectation value $\langle\hat{A}\rangle$ of a measured
quantum observable $\hat{A}$ is given by the inner product formed between the
current quantum state $\vert\Psi\rangle$ of the physical system, and the transformed quantum state when acted upon by the observable, $\hat{A}\vert\Psi\rangle$, that is
\begin{equation*}
\langle\hat{A}\rangle=\langle\Psi\vert\hat{A}\vert\Psi\rangle
.
\end{equation*}

\paragraph{Q3 (dynamics)}
The time evolution of a closed physical system obeys the Schr\"{o}dinger~equation
\begin{equation*}
\imath\hbar\frac{\partial}{\partial t}|\Psi\rangle=\hat{H}|\Psi\rangle ,
\end{equation*}
where the Hamiltonian $\hat{H}=\hat{H}^{\dagger}$ is Hermitian observable
corresponding to the total energy of the system.
The general solution of the Schr\"{o}dinger equation could be expressed
with the use of the matrix exponential function of the Hamiltonian:
\begin{equation*}
|\Psi(t)\rangle=e^{-\frac{\imath}{\hbar}\hat{H}t}|\Psi(0)\rangle ,
\end{equation*}
where $|\Psi(0)\rangle$ is the initial quantum state at time $t=0$.
%
%



\end{document}